# Automatic segmentation of clear cell renal cell tumors, kidney, and cysts in patients with von Hippel-Lindau syndrome using U-net architecture on magnetic resonance images

Pouria Yazdian Anari[1], Nathan Lay[2], Aditi Chaurasia[1], Nikhil Gopal[3], Safa Samimi[1], Stephanie Harmon[2], Rabindra Gautam[3], Kevin Ma[2], Fatemeh Dehghani Firouzabadi[1], Evrim Turkbey[1], Maria Merino[4], Elizabeth C. Jones[1], Mark W. Ball[3], W. Marston Linehan[3], Baris Turkbey[2], Ashkan A. Malayeri[1*]

1. Radiology and Imaging Sciences, Clinical Center, National Institutes of Health, USA.

2. Artificial Intelligence Resource, National Institutes of Health, USA.

3. Urology Oncology Branch, National cancer institutes, National Institutes of Health, USA.

4. Pathology Department, National Cancer Institutes, National Institutes of Health, USA.

**Corresponding author:**

Ashkan A. Malayeri, MD

**Address:**

Clinical Center, National Institutes of Health

10 Center Drive, 1C352

Bethesda, Maryland 20892

Email: ashkan.malayeri@nih.gov

Phone: (301) 451-4368

**Abbreviations:**

**ccRCC**: clear cell renal carcinoma

**VHL**: Von Hippel-Lindau

**CNN**: Convolutional neural networks

**DSC**: Dice Similarity Coefficient

**AUC**: Area under the curve

**MRI**: Magnetic resonance imaging

**Abstract**

We demonstrate automated segmentation of clear cell renal cell carcinomas (ccRCC), cysts, and surrounding normal kidney parenchyma in patients with von Hippel-Lindau (VHL) syndrome using convolutional neural networks (CNN) on Magnetic Resonance Imaging (MRI). We queried 115 VHL patients and 117 scans (3 patients have two separate scans) with 504 ccRCCs and 1171 cysts from 2015 to 2021. Lesions were manually segmented on T1 excretory phase, co-registered on all contrast-enhanced T1 sequences and used to train 2D and 3D U-Net. The U-Net performance was evaluated on 10 randomized splits of the cohort. The models were evaluated using the dice similarity coefficient (DSC). Our 2D U-Net achieved an average ccRCC lesion detection Area under the curve (AUC) of 0.88 and DSC scores of 0.78, 0.40, and 0.46 for segmentation of the kidney, cysts, and tumors, respectively. Our 3D U-Net achieved an average ccRCC lesion detection AUC of 0.79 and DSC scores of 0.67, 0.32, and 0.34 for kidney, cysts, and tumors, respectively. We demonstrated good detection and moderate segmentation results using U-Net for ccRCC on MRI. Automatic detection and segmentation of normal renal parenchyma, cysts, and masses may assist radiologists in quantifying the burden of disease in patients with VHL.

**Introduction**

The von Hippel-Lindau (VHL) syndrome is an inherited disease that manifests itself in a variety of lesions across the body, including clear cell renal cell carcinoma (ccRCC)[1, 2]. Renal cell carcinoma is the leading cause of mortality in VHL patients and can range from a complex cyst to a solid tumor[3]. Patients generally develop bilateral, multifocal renal tumors, often requiring multiple surgeries throughout their lifetime. MRI is the imaging method of choice for lifelong surveillance due to its lack of radiation. On the other hand, up to 90% of sporadic nonfamilial ccRCC lesions have been shown to have a VHL pathological variation[4, 5, 6].

While any renal mass with a solid component is assumed to be malignant in VHL patients, additional characterization of a tumor (i.e., growth rate or grade) is presently unable to be assessed on a single imaging study. Radiomics has emerged as a powerful tool to noninvasively predict tumor biological behavior and enhance the ability to individualize treatment for patients. However, radiomic analysis traditionally relies on manual segmentation of tumors, which is a laborious task, particularly in VHL kidneys where numerous cysts and/or post-operative changes from multiple prior partial nephrectomies limit the ability to identify renal tumors. In order to improve the efficiency of radiomics workflow, such that it can be eventually implemented in a clinical setting, automated segmentation of renal lesions is warranted. Almost all recent research on automated RCC detection and segmentation utilizes computed tomography (CT) scans owing to the ubiquity of CT imaging for renal mass assessment and public availability of the MICCAI 2019 Kidney and Kidney Tumor Segmentation (KITS19) data set[7, 8]. KITS19 is a CT RCC grand challenge data set with the objective being to segment kidney and RCC tumors. The top-scoring method of KITS19 had a tumor Dice Similarity Coefficient (DSC) of 0.86. The vast majority (if not all) of recent renal mass segmentation methods are based on deep convolutional neural networks (CNN) (i.e., variations of 2D and 3D U-Net such as nnU-Net)[9].

To our knowledge, no published study has built a deep learning model based on MRI scans of VHL patients, and no current publication on MRI has examined the auto-segmentation of parenchyma, cysts, and tumors simultaneously in the kidney. In this study, our purpose was to demonstrate the application of 2D and 3D U-Net to automatically segment all components of the kidney (e.g., normal parenchyma, cysts, and tumors) in VHL patients from contrast-based MRI.

**Material and Methods**

**Patient cohort**

A registry of VHL patients who underwent renal surgery between January 2015 and June 2021 were queried. Patients had been enrolled on an IRB-approved clinical protocol, with written consent obtained. The research was conducted in conformity with the ethical guidelines outlined in the Declaration of Helsinki in 1964 and its subsidiary legislation. Patients included in this study had clear cell renal cell carcinoma. MRIs were captured before the surgery with the average duration from MRI to surgery being 77.32 days (minimum of 1 day and maximum of 1086 days). Patients with no or low-quality MRIs were excluded from the study (**Figure 1**). MRI examinations were conducted on four MRI scanners. Additional technical information about the MRI scanners can be found in supplementary table 1. Patients underwent the following routine imaging sequences: multiplanar T2, pre-contrast T1, and post-contrast T1. One dose of gadobutrol (0.1 mmol per kilogram of body weight, Gadovist, Bayer, Washington, DC) was administered followed by a 20 mL saline flush. After contrast material injection, images were acquired on axial plane during the (20 second), nephrogenic (70 second), and excretory (3 minute) phases.

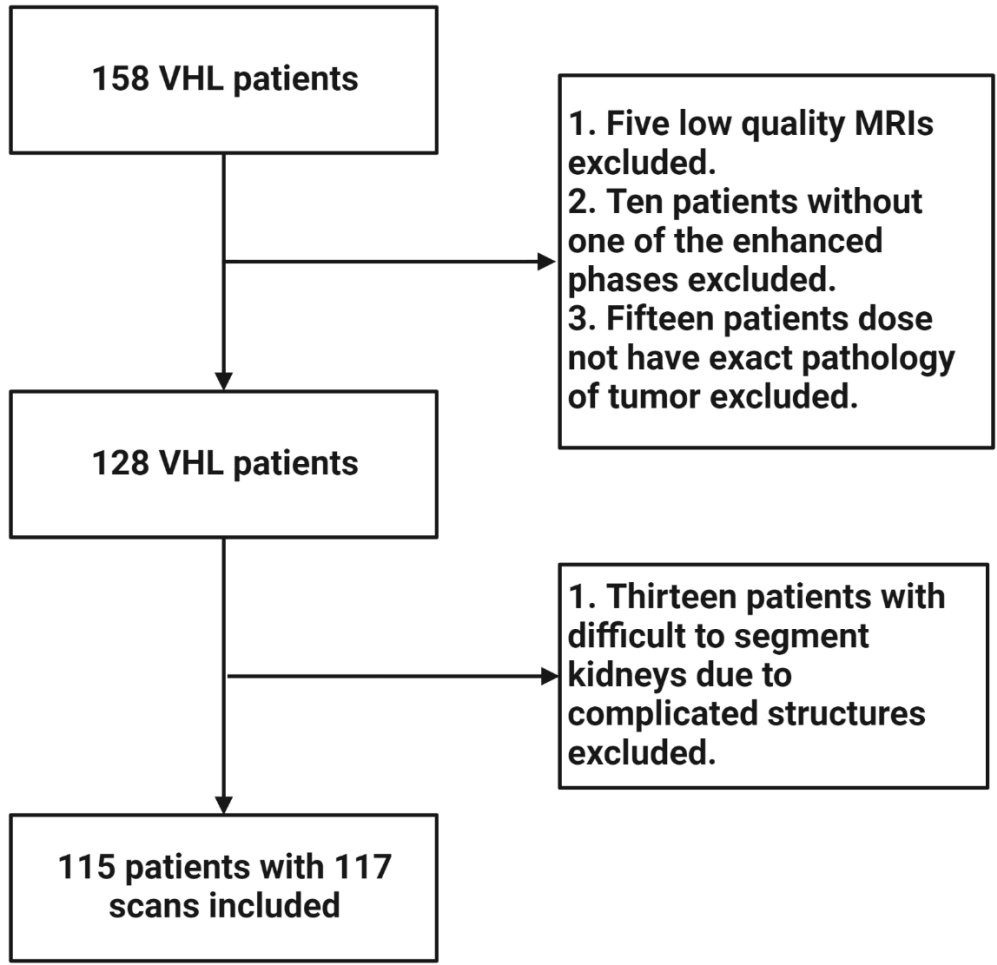

**Figure 1. Diagram of participants who were excluded from the study and those who were included.** From 158 patients we included 115 images with 2 patients having both kidneys annotated (117 halved images) with 504 lesions.

Two postdoctoral radiology research fellows with 1 year of experience in reading MRI segmented the kidney parenchyma, tumors, and cysts using ITK-SNAP (version 3.8) on axial plane of excretory (3 minute) phases. All segmentations were reviewed by an MRI fellowship-trained body radiologist (Ashkan A. Malayeri, 10 years of experience) (**Figure 2**).

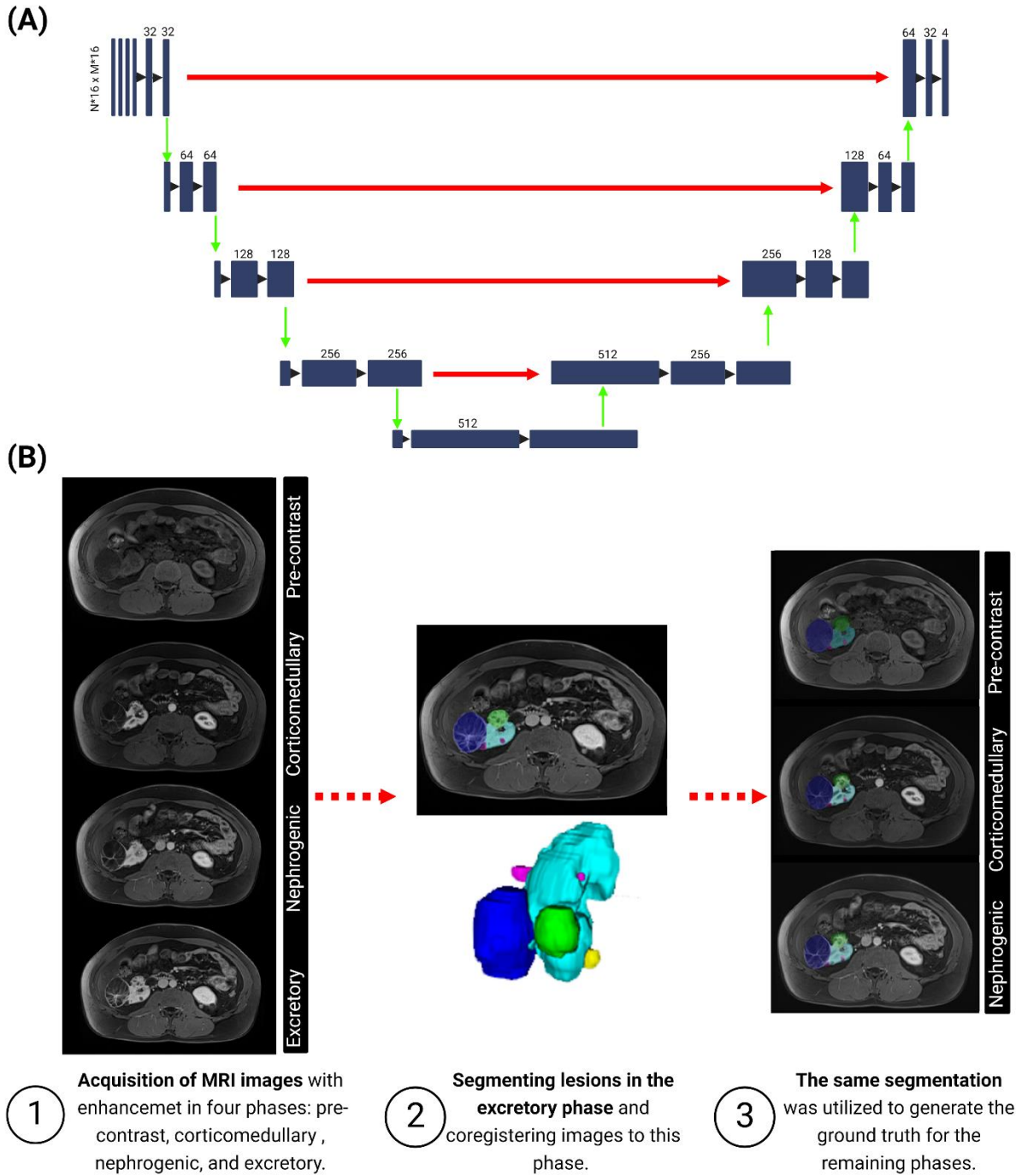

**Figure 2. Workflow diagram of segmentation and architecture of both 2d and 3d U-Net.**
(A) Two-dimensional U-network architecture. (B) The workflow for creating ground truth begins with image acquisition and continues with segmentation and co-registration.

The kidney with the pathology was segmented. It might be one or both kidneys. Each kidney segmentation is saved independently. In cases of innumerable cysts, as often seen in cases with VHL, manual segmentation was performed for all large cysts, but smaller cysts such as those < 10mm remained unsegmented, which were subsequently identified using pixel density and excluded (

**Figure 3**).

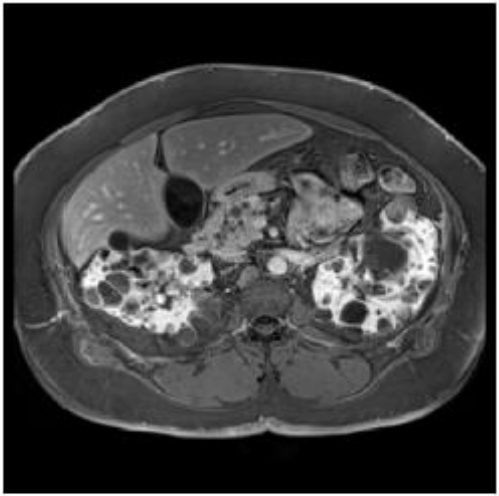 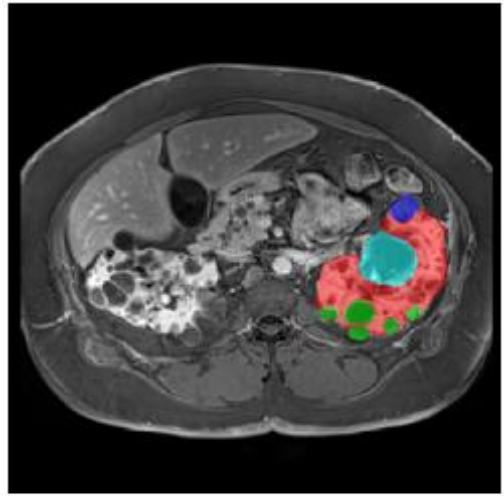

(a) Post-contrast (3 minutes).      (b) Ground truth mask.

**Figure 3. Example of numerous unannotated cysts in ground truth mask.** Segmenting all cysts would be prohibitively laborious. The red, green, blue and cyan are kidney, cyst, tumor, and unknown lesion, respectively.

### U-Net

A U-Net was trained to segment the kidney, cysts, and RCC tumors in 2D axial slices [10]. The U-Net in this work had some additional modifications, including using leaky ReLU (negative slope = 0.2) instead of ReLU[11]. Batch normalizations were inserted before all activations. Renal masses with unknown pathology were excluded from the learning task using an "ignore label."

In other words, regions labeled with the *ignore label* did not contribute to the loss or gradient calculations. Using intensity thresholding, hypointense regions inside the kidney were masked out with the same *ignore label* to deal with the numerous unannotated cysts.

The U-Net was initialized with random weights and trained to minimize binary cross-entropy using Adam stochastic gradient descent from Kingma et al. (Learning rate = 0.001, α = 0.001, $\beta_1 = 0.9$ and $\beta_2 = 0.999$)[12]. As many annotations only featured a single annotated kidney, the U-Net was trained to segment on sagittal-halved images. Horizontal image flipping was used for data augmentation.

The 3D U-Net was prepared identically as the 2D U-Net but used 32x32x32 patches owing to performance and GPU memory constraints. Training considered minibatches of 16 32x32x32 patches with a stride of 16x16x16. Inference evaluated 32x32x32 patches with a stride of 28x28x28. Overlapping probability values were aggregated by taking the maximum value.

Average dimension of images for 3D U-Net was X = 378.8 ± 34.0, Y = 330.9 ± 39.4 and mean slice number of 90.4 ± 5.8, median size of X = 380.0, Y = 326.0, Z = 88.0.

### Data Preprocessing

From each MR scan, the T1 pre-contrast, 20 second post-contrast, 70 second post-contrast, and 3-minute post-contrast images were identified, registered, and finally resampled to have 1 mm x 1 mm axial slices. For 3D U-Net models, images were resampled to 1mm x 1mm x 3mm while slice spacing remained the same for 2D U-Net models. Images were registered using niftyreg (footnote with URL) which uses a combination of rigid and affine image registration[13] followed by fast free-form deformation[14]. All MRIs were normalized using Rician normalization[15]. Examples of these MRIs and ground truth segmentations can be found in **Figure 2**.

**Statistical Analysis**

The U-Net performance was evaluated on ten randomized splits of the patients. Each split used 75% for training and 25% for testing. Ten percent of the training images were used for validation, where the dice similarity coefficient (DSC) was used to pick the best model state (every 50 epochs). DSC is defined as

$$\text{DSC}(S_{\text{unet}}, S_{\text{gt}}) = \frac{2|S_{\text{unet}} \cap S_{\text{gt}}|}{|S_{\text{unet}}| + |S_{\text{gt}}|}$$

where |.| is the cardinality of a set and $S_{\text{unet}}$ and $S_{\text{gt}}$ are sets of voxel positions contained in the predicted segmentation and ground truth segmentation, respectively.

To ensure training and test sets were representative of the full population, each randomized split was evenly sampled from a list of easy and complex cases. A case with 10 or more annotated cysts was considered complex. This gave a total of 65 easy cases and 52 complex cases.

When evaluating DSC, all tumors were treated as one mask. This can cause bias toward the DSC of larger tumors. For example, if a large tumor had a high DSC and smaller tumors had a lower DSC, the overall DSC would be high. For this reason, we additionally measured the average per-tumor DSC by calculating the connected components of the predicted tumor mask and mapping them to ground truth tumors. Then we measured the DSC of overlapping connected components to get the DSC for each ground truth tumor. These were averaged per patient and then averaged over all patients to get the average per-tumor DSC.

In addition to DSC, AUC of ccRCC tumor detection was also evaluated. For a given probability threshold, a tumor was considered detected if the 90$^{\text{th}}$ percentile of the probabilities inside the tumor exceeded the threshold. This means that at least 10% of the tumors had probability scores exceeding the threshold. False positives were calculated in a similar fashion in a 5 x 5 x 5 mm grid inside the kidney, but excluding cyst, tumors, and other renal masses.

For calculating standard deviation for AUC we used the following equition:

$$\bar{f}(x) = \frac{1}{N}\sum_{n=1}^{N} f_n(x)$$

$$\text{std}\left(\{f_n\}_{n=1}^{N}\right) = \sqrt{\frac{1}{N-1}\sum_{n=1}^{N}\int_0^1 (f_n(x)-\bar{f}(x))^2 dx}$$

$$= \sqrt{\int_0^1 \frac{1}{N-1}\sum_{n=1}^{N}(f_n(x)-\bar{f}(x))^2 dx}$$

**Computing Environment**

All experiments and analysis were run on a Linux (CentOS 7.5) computer equipped with an Intel Xeon Gold 6140 processor and NVIDIA Tesla V100x GPU and running Python 3.8 and PyTorch 1.10 configured for CUDA 10.2 and cuDNN 7.6.

**Results**

This research used an image database of 115 individuals and 117 scans with 504 pathologically confirmed ccRCCs and 1171 cystic lesions. The mean number of ccRCC lesions per patient was $4.3 \pm 3.1$ (Table 1).

**Table 1 Demographic characteristics of the patients included in the study.**

| Variable | | Number (%) |
|---|---|---|
| Gender | Male | 65 (56.5) |
| | Female | 50 (44.4) |
| Tumor grade | Fuhrman grade 2 | 424 (84.1) |

|  | **Fuhrman grade 3** | 73 (14.4) |
| --- | --- | --- |
|  | **Fuhrman grade 4** | 17 (2.5) |
|  | colspan="2" Mean (min to max) | |
| **Age** | colspan="2" 48.7 (23–78) | |
| **Tumor per patient** | colspan="2" 4.3 (1–17) | |
| **Segmented cyst per patient** | colspan="2" 10.1 (1–21) | |

The average performances over the 10 randomized splits are shown on Table 2. **Figure 4** show examples of good and bad ccRCC segmentations. While the lesion segmentations are relatively low on average, the U-Net was able to detect tumors with relatively high AUC (0.89) even in the worst case (i.e., >10 cysts in a kidney).

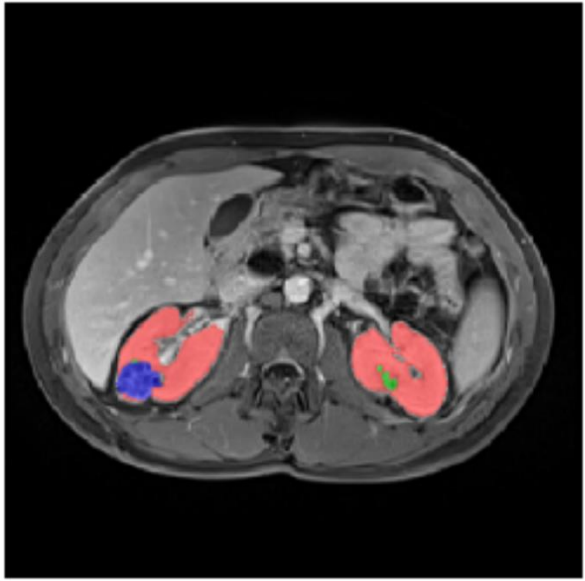
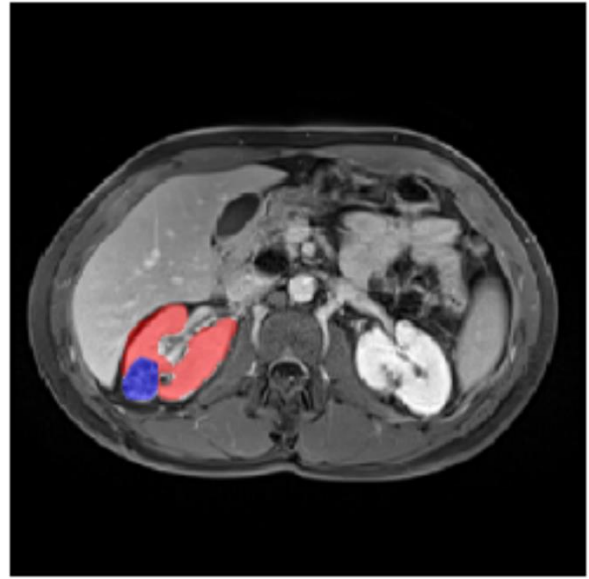

A. Best U-Net prediction — Ground truth

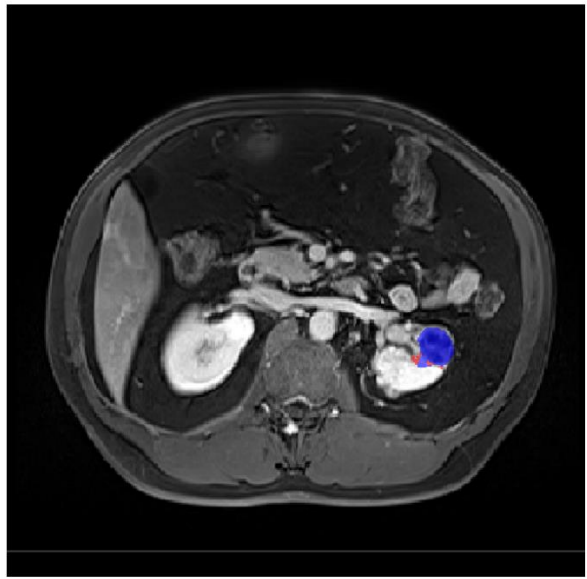
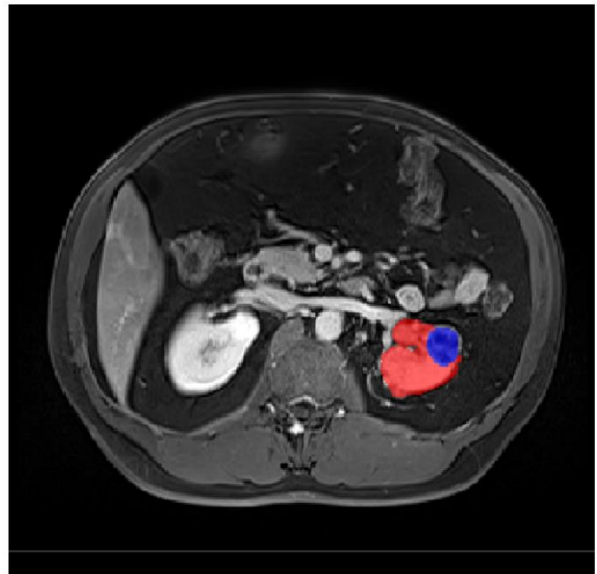

B. Worst U-Net prediction — Ground truth

**Figure 4. Examples of the best and worst U-Net test set segmentations and ground truths.** These are the outputs of the best performing model from the best performing random split experiment. The red, green and blue colors correspond to kidney, cyst and ccRCC tumor. **A.** Dice score of best performance of the best model for kidney = 0.89, Cyst = 0.68, and Tumor = 0.62. **B.** Dice score of worst performance of the best model for kidney = 0.21, Cyst = 0.38, and tumor = 0.0

**Table 2 2D and 3D U-Net average performance.** Results for kidney, cyst, tumor, and per tumor DSC (DSC [Standard Deviation]) and tumor detection area under the curve (AUC).

|  | **Kidney** | **Cyst** | **Tumor** | **Per-Tumor** | **AUC** |
|---|---|---|---|---|---|
| **U-Net 2D** | 0.78 (0.15) | 0.40 (0.24) | 0.46 (0.21) | 0.31 +/- 0.27 | 0.88 |
| **U-Net 3D** | 0.67 (0.20) | 0.32 (0.23) | 0.34 (0.23) | 0.23 +/- 0.26 | 0.79 |

As illustrated in Figure 5, the model accurately detected nearly 100% of the background in all splits. Frequently, kidneys are misclassified as background, cysts as background, and tumors as either kidney or background.

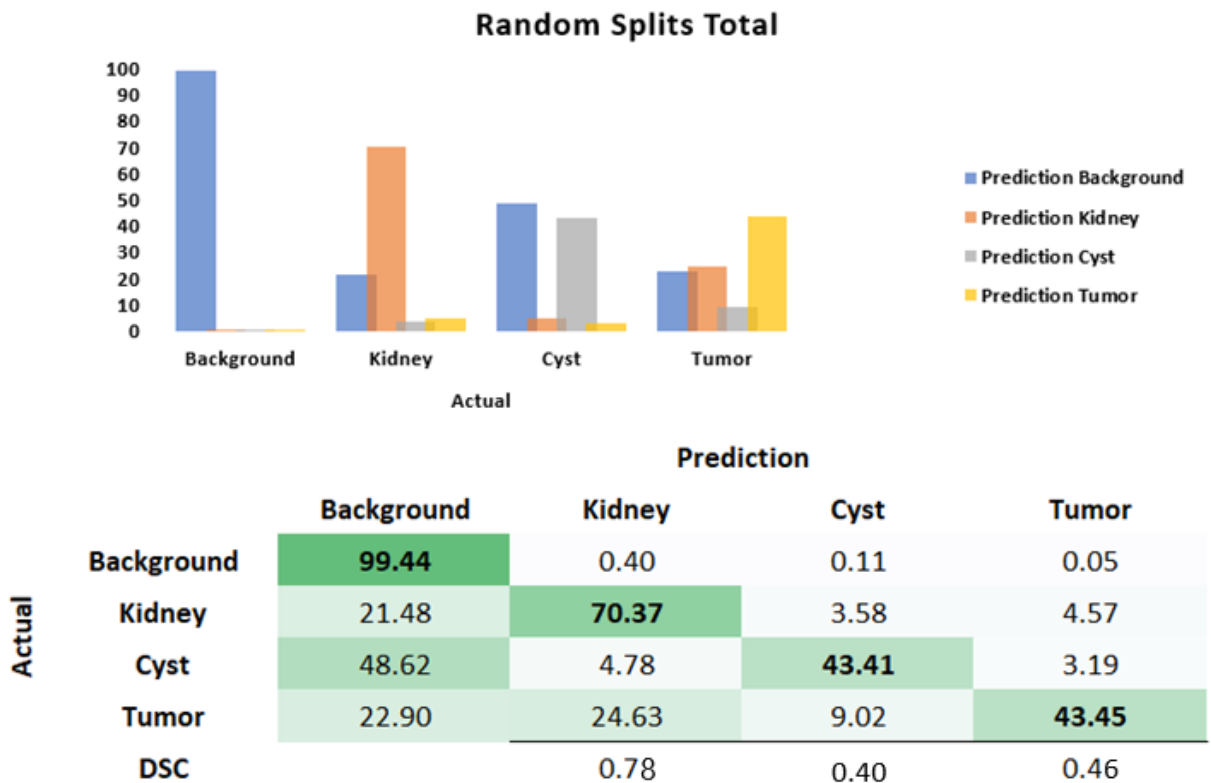

**Figure 5. The voxel percent confusion matrix and box plot illustrate the misclassification of each label in the 2D U-net.** Tumor pixels are mostly detected as kidney or background falsely.

As the ROC curve demonstrates, the area under the curve of 3D-U-Net performance was lower than 2D U-Net, but there was no significant difference between these two models' performances on detecting the ccRCC lesions (Figure 6).

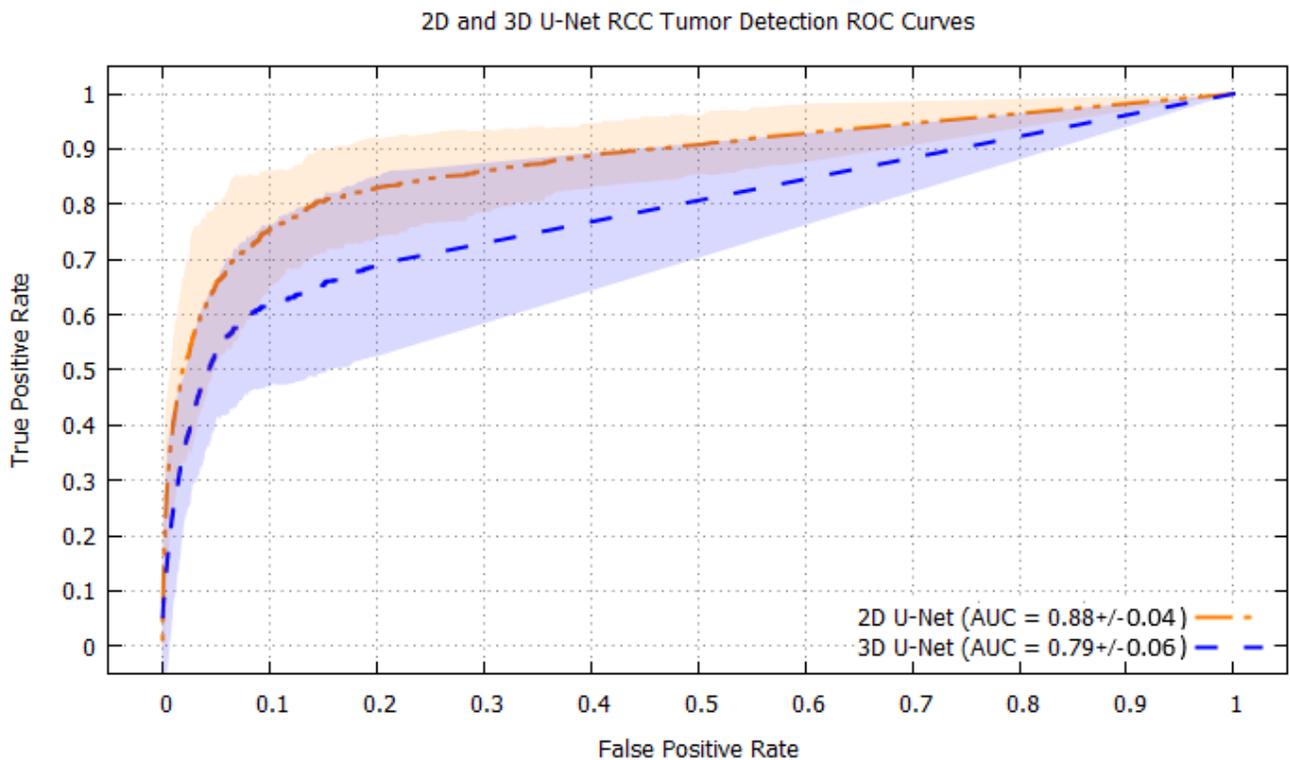

**Figure 6. Tumor Detection ROC curve.** The yellow and blue shaded regions reflect two standard deviations of detection rate for a given false positive rate.

**Discussion**

This study aimed to determine if CNN (2D and 3D U-Net) could be used to automatically identify and segment ccRCC tumors from cysts and the surrounding renal parenchyma in VHL patients. Our algorithm had high accuracy in detecting renal tumors on MRI (AUC 0.79–0.88) and showed promise in the segmentation of these tumors (DSC 0.34–0.46). VHL kidneys are often comprised of numerous cysts. Locating tumors in these types of kidneys can be akin to finding a needle in a haystack; however, as we have demonstrated, artificial intelligence may help orient radiologists to these lesions, allowing for prompt and appropriate management to avoid the morbidity of metastatic disease from missed lesions. While automatic detection and segmentation of RCC tumors has been performed with CT, prior to this study, no such endeavors have been performed using MRI[7]. With its lack of radiation and its improvement in tissue characterization, MRI increasingly utilized to assess renal lesions. As such, developing auto-segmentation algorithms on MRI to facilitate tumor detection in the clinical environment and workflow for future radiomic studies is warranted.

**Comparing the performance**

By examining prior studies on CT scans, we can observe that a continual learning curve was employed to bring automatic segmentation of renal cell carcinoma to its current stage. KITs-19 signaled the start of a new chapter in this mission[7, 16]. The majority of models in the KITs-19 challenge demonstrated encouraging performance in terms of tumor detection and renal parenchyma detection. Although our study's DSC scores for tumor segmentation are not as good as those in comparable CT-based articles, we lay the groundwork for future MRI auto-segmentation models Additionally, MRI imaging has several challenges compared to CT, mainly the duration of scans and the effect of breathing on image quality at spatial registration. Breathing time shifts make image co-registration challenging.

At the time of writing this article, no other peer-reviewed research has attempted to segment ccRCC lesions and cysts using either 2D or 3D U-Net, either in VHL patients or on MRI. The only comparable research to ours is the thesis of Agarwal. et al., in which they employed a similar approach for automatically segmenting the different types of RCC tumors on the nephrogenic phase[16]. Our model demonstrated an improvement in tumor segmentation compared to mentioned study. It can be due to the homogeneity of our cohort of VHL patients. for this matter, we need to evaluate this model on sporadic ccRCC patients as well.

Regarding auto-segmentation of kidney parenchyma, Agarwal et al., Taro Lagner et al., and the CHAOS challenge had DSC scores of more than 90% in their population of healthy volunteers. All three studies had DSC scores for kidney parenchyma better than our model[17, 18]. Results for kidney parenchyma from our method may differ from the CHAOS and KITS-19 Challenge sets due to differences in ground truth definition of "kidney" regions, for which both challenges defined foreground as kidney parencheyma and all structures contained within kidneys (tumors, cysts). Our results are reported for foreground label as exclusively normal kidney parenchyma. However, as previously mentioned, the kidney parenchyma in VHL patients can be altered due to surrounding multifocal cystic and solid lesions as well as postoperative changes, making segmentation more challenging than in a sporadic RCC or control population. Nevertheless, the model proposed in this work achieved a defensible DSC.

Patients with VHL may develop many cystic lesions in their kidneys that can be simple or complex, with varying times containing hemorrhagic or proteinaceous products. An analogous population with bilateral and/or multifocal cysts are those with polycystic kidney disease (PKD). Kline, et al. and Mu, et al., in two separate studies, reported a DSC for segmentation of renal cysts that was better than our study[19, 20]. Different signal intensities of VHL cysts can complicate machine detection and segmentation of cysts. The range in cystic forms and morphologies, and the existence of numerous solid cystic lesions pathologically confirmed to be tumors (unlike PKD, complex cysts in VHL are often malignant), may account for the lower DSC found in our study.

Another challenging aspect to detection and segmentation of structures in the kidney in our work was the presence of a many unannotated structures. Often, there are dozens of cysts (some very small) that are impossible to annotate. Some pathologies of suspicious masses are also unknown, as they were not excised. If U-Net properly segments unannotated structures in the kidney, the DSC performance for kidney, cyst and/or tumor can suffer. In this case, our model's accuracy in lesion segmentation would be higher than what is reflected in the DSC score. To overcome unlabeled cysts during training, T1 hypointense regions that are labeled kidney are relabeled with an "ignore label." The "ignore label" will ensure unlabeled cysts have no contribution to the loss calculation or gradient updates in training. Tumors with unknown pathologies are similarly relabeled with the "ignore label." In both cases, they neither directly help nor harm training. Consequently, any benign renal structure that is neither tumor nor cyst will not be learned and might harm model's generalizability.

Overall, 3D U-Net performed worse than 2D U-net, which was unexpected. The performance gap between 2D and 3D U-Net is most likely due to a few factors, including whole slice compared with localized 3D patch; sampling strategy of 3D patches; 3D patch size; and cropping effects of the 3D patches. Due to GPU memory constraints, the 2D U-Net can access more global image context while the 3D U-Net is more localized. The work of Isensee proposes automated hyperparameter tuning of both 2D and 3D U-Net for more optimal performance[21]. While we did not implement the methodology of Isensee, we also considered 64x64x64 patches and surprisingly found worse overall tumor DSC.

**Limitation**

Our study has some limitations that can be addressed in future research to enhance the model's performance in automatically segmenting lesions. The first limitation may be the study's small sample size. We identified some under-represented tumors that degraded the model's performance. These underrepresented tumors may be detected by the presence of additional lesions from additional participants.

As the second limitation, this model was built on just four contrasts enhanced T1 sequences; improved performance may be expected with the addition of other MR sequences such as T2 and DWI. However, the most challenging part of adding these sequences will be the co-registration of the images. This problem can be overcome by segmenting the lesions on these sequences alone and contrast-enhanced T1 sequences or by improving the available co-registration software.

**Conclusion**

We demonstrated some encouraging findings for 2D U-Net and 3D U-Net detection and segmentation of ccRCC tumors on MRI for VHL patients. Additionally, this topic necessitates meticulous data set development owing to the wide variety of pathologies and appearances. When the training set is representative, U-Net generalizes better, albeit the definition of representative is not yet apparent.

**Compliance and Ethical Standards**

This single-institution study was approved by the institutional review board. Written informed consent was obtained from all patients.

**Acknowledgments**

This work utilized the computational resources of the NIH HPC Biowulf cluster. (http://hpc.nih.gov). The authors thank Yolanda L. Jones, NIH Library, for editing assistance. We are thankful to the Office of Biomedical Translational Research Informatics (BTRIS), particularly Gloria Oshegbo, for their assistance and provision of necessary data. Additionally, we must